\documentclass[12pt,preprint]{aastex}

\usepackage{graphicx}
\begin{document}

\title{Globular Cluster Formation at High Density:
A model for Elemental Enrichment with Fast Recycling of Massive-Star Debris}

\author{Bruce G. Elmegreen\altaffilmark{1}}

\altaffiltext{1}{IBM Research Division, T.J. Watson Research Center, 1101 Kitchawan Road,
Yorktown Heights, NY 10598; bge@us.ibm.com}

\begin{abstract}
The self-enrichment of massive star clusters by p-processed elements is shown to
increase significantly with increasing gas density as a result of enhanced star
formation rates and stellar scatterings compared to the lifetime of a massive star.
Considering the type of cloud core where a globular cluster might have formed, we
follow the evolution and enrichment of the gas and the time dependence of stellar
mass. A key assumption is that interactions between massive stars are important at
high density, including interactions between massive stars and massive star
binaries that can shred stellar envelopes. Massive-star interactions should also
scatter low-mass stars out of the cluster. Reasonable agreement with the
observations is obtained for a cloud core mass of $\sim4\times10^6\;M_\odot$ and a
density of $\sim2\times10^6$ cm$^{-3}$. The results depend primarily on a few
dimensionless parameters, including, most importantly, the ratio of the gas
consumption time to the lifetime of a massive star, which has to be low,
$\sim10$\%, and the efficiency of scattering low-mass stars per unit dynamical
time, which has to be relatively large, such as a few percent. Also for these
conditions, the velocity dispersions of embedded globular clusters should be
comparable to the high gas dispersions of galaxies at that time, so that stellar
ejection by multi-star interactions could cause low-mass stars to leave a dwarf
galaxy host altogether. This could solve the problem of missing first-generation
stars in the halos of Fornax and WLM.
\end{abstract}


\section{Introduction}
\label{intro}

Most globular clusters (GCs) in the Milky Way have two populations of stars in
approximately equal proportion with a first generation (G1) relatively abundant in
Oxygen compared to Sodium and a second generation (G2) the reverse. This bimodality
is presumably the result of contamination of the G2 stars by the products of
p-process nucleosynthesis in the G1 stars, particularly the reactions $^{22}$Ne to
$^{23}$Na along with the simultaneous destruction of $^{16}$O at a temperature of
$2\times10^7$ K \citep{den90, dec07a}. Other elemental anti-correlations include Mg
with Al \citep[e.g.,][]{carretta14}, explained by \cite{langer93} and others as a
result of high temperature ($T>7\times10^7$ K) proton-capture plus beta-decay that
transforms $^{24}$Mg into $^{25}$Mg, $^{26}$Mg and $^{27}$Al. Also in some GCs,
stellar Nitrogen anti-correlates with Carbon and Oxygen \citep{dickens91}, with a
total C+N+O abundance that is about constant, suggesting a CNO cycle. Reviews of
these abundance anomalies are in \cite{gratton04}, \cite{char05}, \cite{gratton12},
\cite{renzini15}, and \cite{bastian15}.

Light element anomalies from the CNO cycle and the NeNa- and MgAl-chains are
peculiar to the GCs and are not in field or halo stars \citep{gratton00, char05,
prantzos06}. Because they involve high-temperature reactions and the stars in which
they are observed today are too low in mass to have had such p-process reactions
themselves \citep{gratton01}, the Na-excess and Al-excess stars in current GCs had
to form in gas that was pre-enriched with these elements \citep{cottrell81}, most
likely from the previous generation of stars as mentioned above. The debris from
this previous generation may also have mixed with some left-over initial gas to
explain an anticorrelation between Li, which is destroyed in stars, and Na, which
is produced \citep{pasquini05, bonifacio07, dec07a}, and to explain the difference
between the high enrichment in the nuclear burning regions of massive stars and the
observed enrichment in low mass stars \citep[][table 4]{dec07b}.

The first generation has been proposed to include normal massive stars
\citep{cottrell81}, such as rapidly spinning massive stars \citep{prantzos06},
which, because of their rotation, bring p-processed material from the H-burning
zone into the envelope and then shed it along the equator via centrifugal force and
radiation pressure \citep{prantzos06, dec07b}. Binary massive stars with Roche lobe
overflow are also an option \citep{demink09}. AGB stars \citep{dercole08} shed
processed gas at low speed too and may be a source of the anomalies, although the
near-constant C+N+O does not look like a byproduct of AGB stars, which make carbon
during Helium burning \citep{char05}; NGC 1851 may be an exception though
\citep{yong14,simpson16}.  \cite{renzini15} discuss ways in which the AGB option
might still be viable. Other models for the bimodality include proto-stellar disk
accretion \citep{bastian13a}, supermassive stars \citep{denis14,denis15},
GC-merging in dwarf galaxy hosts \citep{bekki16}, and AGB wind re-collection in the
pressurized cavity around the GC \citep{dercole16}.

% paren after simpson16

A problem with these models is that the stellar debris is only a small fraction of
the total stellar mass in a normal IMF. This implies that because the current G1
and G2 masses are comparable to each other, the mass in the original G1 population
had to be at least the inverse of this fraction times the current G1 mass. For the
model with rapidly rotating massive stars and a normal IMF, the original cluster
had to be $\sim20$ times the current G1 mass \citep{dec07b}, and for the AGB model,
the original cluster had to be $>10$ times the current mass \citep{dercole08}.  The
missing 90\%-95\% of the G1 stars had to escape, and even more had to escape if G2
stars escaped too, which is likely.  These escaping stars presumably comprise the
halo of the Milky Way \citep{prantzos06,martell11}, but such large halo amounts are
not evident in the Fornax or WLM dwarf galaxies
\citep{larsen12,larsen14,elmegreen12}.

Another constraint on the models is that there is virtually no elemental
contamination from G1 supernova inside the G2 stars \citep{char05,renzini13}. In
the model with rapidly rotating massive stars, this constraint implies that the G2
stars formed quickly, before the G1 supernovae exploded. A potential problem here
is that without supernovae, the gas that formed G1 may not be cleared away from a
massive cluster \citep{ginsburg16,lin16}, causing a high fraction of it to turn
into stars. Such high efficiency prevents the cluster from shedding its excess G1
stars during rapid gas loss \citep{krause16}, and then the final cluster is too
massive and it has too many G1 stars compared to G2. Cluster clearing was tested by
\cite{bastian14b}, who found that clusters with masses up to $10^6\;M_\odot$ and
ages between 20 and 30 Myr are often clear of gas. This age is old enough for some
supernovae to have occurred. Without gas, star formation stops and a prolonged
epoch of secondary star formation that feeds on G1 debris does not occur.

Secondary or delayed star formation as in the AGB model is also not observed in
today's massive clusters at the predicted age range of 10-1000 Myr
\citep{bastian13b,cabrera14}, although it was reported by \cite{li16} whose result
was then questioned by \cite{cabrera16a}. Neither do massive 100 Myr old clusters
have obvious gas from accumulated winds or secondary accretion \citep{bastian14a}.

Here we investigate further the model by \cite{prantzos06} and others, where
p-process contamination from massive stars is quickly injected into a star-forming
cloud core and incorporated into other forming stars. We consider conditions
appropriate for the early Universe where the interstellar pressure was much higher
than it is today. This pressure follows from the observation that star formation
was $\sim10$ times more active per unit area than it is now \citep{genzel10} as a
result of a factor of $\sim10$ higher gas column density \citep{tacconi10,daddi10}.
Because interstellar pressure scales with the square of the column density ($P\sim
G\Sigma^2$), the pressure was $\sim100\times$ larger when GCs formed than it is in
typical star-forming regions today. Consequently, the core of the GC-forming cloud
was likely to have a higher density and therefore a larger number of free-fall
times, such as 10 or more, before the first supernovae occurred. Star formation
could have occurred so quickly and the energy dissipation rate at high density
could have been so high that feedback from massive stars had little effect on cloud
dispersal before the supernova era \citep[e.g.,][]{wunsch15}. Also at high density,
massive stars experience a significant drag force from gas and low-mass stars
\citep{ostriker99}, causing them to spiral into an even more compact configuration.
High stellar densities lead to the dispersal of extrusion disks filled with stellar
envelope material \citep{prantzos06}, and also to close encounters between massive
stars or between tight massive binaries and massive stars, which can shred the
stellar envelopes \citep{gaburov10}. High stellar densities may also make
supermassive stars by coalescence \citep{ebisuzaki01,bally05}, and these stars can
produce the highest-temperature p-process elements \citep{denis15}.

A related point is that at high density, a significant fraction of early-forming
low-mass stars should have been ejected from the cloud core by interactions with
massive binaries and massive multi-body collisional systems. This could occur long
before ``infant mortality,'' when the final clearing of gas signals the end of the
star formation process. Rapid and continuous stellar ejections by massive star
interactions are well observed in numerical simulations of cluster formation
\citep{reipurth01,bate05,fujii13}. Because massive three-body collisions can also
dump nearly a supernova's worth of kinetic energy into a region
\citep{gaburov10,umbreit08}, the gas in the core should be continuously agitated.
The corresponding changes in the gravitational potential energy of the gas should
eject even more low-mass stars, in analogy to the proposed ejection of stars and
dark matter from the cores of young star-bursting dwarf galaxies
\citep{governato12,elbadry16}.

Dense cores are likely to continue accreting from the cloud envelope for many core
dynamical times. If the core plus envelope gas mixes with stellar debris and forms
new stars, then a succession of stellar populations will occur with ever-increasing
levels of p-process elements. By the time the first massive stars begin to
supernova and clear away the gas, some 3-7 Myr after star formation begins
\citep{heger03}, there should be a wide range of p-process elements in the stars
that have formed.  The model presented below shows that this range can reproduce
the observations for certain values of dimensionless parameters.

A key observation is that elemental enrichment in GCs seems to be discrete
\citep{marino11,carretta12,carretta14,renzini15}. Discreteness requires burst-like
contamination, and in the stellar interaction model here, that means intermittent
events of catastrophic massive-star interactions. The interactions occur where the
density is highest, so either the stellar density in the cloud core varies
episodically with a burst of stellar collisions following each high-density phase,
or there are several mass-segregated cores in a proto-GC \citep{mcmillan07,fujii13}
and each has its own burst of intense interactions. Both situations are likely and
could contribute to discrete populations of stars. The temporal density variations
would presumably follow from the time-changing gravitational potential in the
cluster core, as stellar envelope mass is disbursed along with cloud mass through
collisions, and then recollected in the core after a dynamical time because of
self-gravity and background pressure.

% comma before "and each"

The following sections examine this model in more detail. Section \ref{basicmodel}
outlines the basic model of GC formation at high density, section \ref{modeleq}
presents the equations that govern this model, giving some analytical solutions in
section \ref{analyticalsol}, and section \ref{results} shows the results. A
conclusion that highlights the main assumptions of the model and their implications
for high-density cluster formation is in Section \ref{conclu}.

\section{Star Formation in Stellar Debris}
\subsection{Basic Model}
\label{basicmodel}

The basic scale of GC formation considered here involves a molecular cloud core of
mass $\sim4\times10^6\;M_\odot$ in a spherical region of radius $\sim3$ pc. A
larger region of lower-density gas should surround this core, possibly in the form
of filaments or spokes which continuously deliver new gas
\citep{klessen98,myers09}. Perhaps the total mass involved with the proto-GC and
its neighborhood is $\sim10^7\;M_\odot$ or more, as observed for a massive dense
region in the Antenna galaxy \citep{herrera12,johnson15}. The core molecular
density for the above numbers is $2.2\times10^6$ cm$^{-3}$ and the free fall time
($=(3\pi/32 G \rho)^{0.5}$) is 0.03 Myr. The ratio of the core mass to the free
fall time, $133\;M_\odot$ yr$^{-1}$, is a measure of the core accretion rate during
core formation. If this core collapsed from the interstellar gas at the typical
rate of $\sigma_{\rm ISM}^3/G$ for interstellar velocity dispersion $\sigma_{\rm
ISM}$, then $\sigma_{\rm ISM}=82$ km s$^{-1}$, which is not unusual for
high-redshift disk galaxies \citep[e.g.,][]{forster09}. The accretion rate of
low-density peripheral gas on to the core should be much less than the initial core
formation rate.

The fiducial cloud core mass of $\sim4\times10^6\;M_\odot$ was chosen to produce a
final GC mass of around $2\times10^5\;M_\odot$, which is at the peak of the GC mass
distribution function \citep{harris79}.  The first factor of $\sim10$ in mass
reduction follows from our model including multiple generations of star formation
in the core and mass loss from stellar ejection, as required by the observed spread
in p-processed elemental abundance. Another factor of $\sim2$ reduction in the GC
mass is likely from evaporation over a Hubble time \citep{mclaughlin08}.

The average core surface density in this basic model, $1.4\times10^5\;M_\odot$
pc$^{-2}$, is comparable to the maximum for stellar systems
\citep{hopkins10,walker16}.  The core velocity dispersion is $\sim76$ km s$^{-1}$,
which is a factor of 2 higher than for massive clusters today, but not unreasonable
for a young galaxy where the gas turbulence speed is this high. Over time, the
cluster should expand and the dispersion decrease \citep[e.g.][]{gieles16}. The
original cluster dispersion is high enough to make feedback-driven gas loss
difficult before the supernova era \citep{matzner15,krause16}. We consider that
because of this difficulty, the efficiency of star formation per unit free-fall
time might be relatively high, $\sim10$\%, instead of the usual 1\%
\citep{krumholz07}. Then the gas consumption time, which is the free-fall time
divided by the efficiency, is 0.3 Myr. The significant point here is that this
consumption time is much less than the evolution time of a high-mass star.

One potential implication of the high velocity dispersion in GC-forming cloud cores
is that stars ejected by time-changing gravitational potentials (e.g., binary or
multi-star interactions and the induced rapid gas motions) should also have fairly
high velocities. Dynamical ejection processes can be much more energetic than
thermal evaporation from a random walk.  This offers an intriguing solution to the
problem stated in the introduction that Fornax and WLM do not have stellar halos
massive enough to include all of the required mass of G1 stars that should have
been ejected from their GCs.  In fact these are dwarf galaxies with very slow
internal motions: the Fornax dwarf spheroidal galaxy has a $\sim12$ km s$^{-1}$
internal velocity dispersion and a much lower rotation speed \citep{walker06},
while WLM has a 36 km s$^{-1}$ rotation speed \citep{leaman12}. Stars that are
ejected at a factor of 1.5 to 2 times the escape speed of the GC could leave the
galaxy. This possibility leads to the prediction that {\it galaxies with slow
internal motions should have a systematic depletion in halo stars from the G1
population that escaped their GCs}.  A similar conclusion was reached by
\cite{khalaj16} based on stellar loss from GCs during gas expulsion.

The IMF for all star formation is assumed to be fully populated and given by the
log-normal distribution in \cite{paresce00} for stellar mass
$0.1\;M_\odot<M<0.8\;M_\odot$, with mass at the peak $M_C= 0.33\;M_\odot$ and
dispersion $\sigma= 0.34\;M_\odot$, and by a power law with the Salpeter slope
$-2.35$ above $0.8\;M_\odot$ \citep[see also][]{prantzos06}.  The upper limit to
the stellar mass will be varied from $M_{\rm upper}=100\;M_\odot$ to
$300\;M_\odot$, with the high value considered because of stellar coalescence. Note
that a $320\;M_\odot$ star has been suggested for the dense massive cluster R136 in
the LMC \citep{crowther10,crowther16}, and stars more massive than $100\;M_\odot$
were found in the dense cluster in NGC 5253 \citep{smith16}. Stars with masses
larger than $20\;M_\odot$ are assumed to make p-process elements \citep{dec07a} and
mix them into their stellar envelopes, which are defined to be all of the stellar
mass outside of the He core, as given by \cite{prantzos06}. For $M_{\rm
upper}=100\;M_\odot$, the fraction of the total stellar mass in the form of these
envelopes is $f_{\rm env}= 7.9$\%, which is the product of the fraction of the IMF
in stars with $M>20\;M_\odot$ (12.1\%) and the average fraction of this stellar
mass in the form of envelopes (65.1\%). For $M_{\rm upper}=300\;M_\odot$, $f_{\rm
env}= 9.3$\% (16.4\% of the IMF is in $M>20\;M_\odot$ stars and 56.8\% of that mass
is in envelopes). Also for these two upper masses, the fraction of the total
stellar mass in long-lived, low-mass stars ($M<0.8\;M_\odot$), is $f_{\rm LM}=
31.2$\% and 29.7\%, respectively. The formation of p-process elements and the
delivery of these elements into the stellar envelopes and equatorial disks is
assumed to proceed at a steady rate with an average timescale $t_{\rm evol}=3$ Myr,
the lifetime of a high-mass star.

Regarding the contamination by p-processed elements, we note that the [O/Na] ratio
in GCs varies from the large value of $\sim 0.4$ in G1 to the small value of $\sim
-1.4$ in G2, depending on the GC \citep[$-1.4$ is for NGC 2802;][]{carretta06,prantzos06,gratton12}.
The maximum value is about the same as in
field stars and the minimum value is close what is expected in the envelope of a
massive star near the end of its pre-main sequence phase \citep{prantzos06}. The
range of observed values, 1.8 dex, is less than the change expected inside the
nuclear burning regions of massive stars, which is 2.8 to 3.4 dex (depending on
reaction rates) in \cite{dec07a}. This difference allows for some dilution of the
core region with the envelope of the star before dispersal. The full range in
today's stars therefore extends from the presumed initial condition when the first
generation formed, to a value that represents the near-complete conversion of a
massive stellar envelope into one or more G2 stars. Values of [O/Na] between these
extremes correspond to some combination of incomplete mixing of the processed
debris with first generation gas, and partially processed gas from stars that have
not yet finished their main sequence evolution \citep[Scenarios II and I
in][]{prantzos06}.

The dilution proportions of processed stellar envelope gas and original cloud gas
has also been estimated from the relative Li abundance in the G2 stars, considering
that Li will be destroyed in the massive G1 stars. The observations suggest that up
to 70\% of the mass of a G2 star could come from the debris of G1 stars
\citep{dec07b}. This high fraction limits the amount of accretion from the cloud
envelope during the star formation process.

As mentioned above, the enriched gas is assumed to come from stellar equatorial
disks \citep{prantzos06} and stellar debris generated by close interactions. For
example, \cite{gaburov10} show that massive binaries that collide with another star
can merge and puff up \citep[see also][]{fregeau04}, and that continued collisions
with the third star before merging can shred the common envelope. Mass loss
fractions of $\sim10$\% are feasible in this situation. \cite{umbreit08} also
consider the implications of collisions and suggest that the kinetic energy from
collision-induced mass loss can clear gas away every few million years from aging
globular clusters. This kinetic energy is comparable to that of a supernova
\citep{gaburov10}.  For the dense cloud cores considered here, that is not enough
energy to clear away the gas, which is highly dissipative, but it could be enough
for later stages of cluster formation after a significant amount of gas has turned
into stars, and that is also when the supernovae themselves begin to clear the
clusters of residual gas.  We assume here that all of the stellar debris, i.e.,
from massive-star equatorial disks \citep{prantzos06}, massive star collisions, and
Roche lobe overflow around massive binaries \citep{demink09}, brings p-processed
material from the envelopes of massive stars into the dense cloud core where it
mixes with existing and newly accreted core gas on the turbulent crossing time
($\sim0.1$ Myr), and is incorporated into new stars.  This process continues for
$\sim3$ Myr, forming the whole final cluster, at which point supernovae begin to
remove the remaining gas.

\subsection{Model Equations}
\label{modeleq}

Consistent with these assumptions, we consider an initial cloud core of mass
$M_{\rm gas}(t=0)$ in which stars begin to form, and a continuous accretion of new
cloud gas onto this core at a rate $R_{\rm acc}$. Stars are assumed to form in the
core with a constant consumption time, $t_{\rm consume}$ (equal to $t_{\rm ff}$
divided by the efficiency per free fall time). The effect of varying the
consumption time will be discussed below. The formation rates of low
($<0.8\;M_\odot$), intermediate ($0.8\;M_\odot-20\;M_\odot$) and high
($>20\;M_\odot$) mass stars are given by the star formation rate in the core
multiplied by the fractions of the IMF in these three mass intervals:
\begin{eqnarray}
dM_{\rm star,LM}/dt= f_{\rm LM}M_{\rm gas}(t)/t_{\rm consume}\label{e1}\\
dM_{\rm star,IM}/dt= f_{\rm IM}M_{\rm gas}(t)/t_{\rm consume}\label{e2}\\
dM_{\rm star,HM}/dt= f_{\rm HM}M_{\rm gas}(t)/t_{\rm consume},
\label{e3}
\end{eqnarray}
where $f_{\rm LM}=0.312$, $f_{\rm IM}=0.567$, and $f_{\rm HM}=0.121$ for an IMF
with a most massive star of $100\;M_\odot$, and where $f_{\rm LM}=0.297$, $f_{\rm
IM}=0.540$, and $f_{\rm HM}=0.164$ for an IMF with a most massive star of
$300\;M_\odot$ (Sect. \ref{basicmodel}). The total mass formed in these stars is
\begin{equation}
M_{\rm star,LM}(t)=\int_0^t {\dot M}_{\rm star,LM} dt, ...
\end{equation}
and so on for the other mass ranges. Here we use the notation ${\dot M}=dM/dt$. The
total mass for all stars is $M_{\rm star}=M_{\rm star,LM}+M_{\rm star,IM}+M_{\rm
star,HM}$.

For the gas, we track the primordial and enriched gas masses separately. The
primordial gas is a combination of what was originally in the cloud core plus what
gets accreted after star formation begins, and it also includes the part of the
stellar envelope debris that was not converted into p-processed elements. We assume
that massive stars make p-process elements continuously and that mixing from rapid
rotation puts these elements into the stellar envelops continuously. Thus we
conceptually divide the envelope mass into an unprocessed fraction at the original
primordial abundance, and a completely processed, or enriched, fraction at the
abundance of fully processed material. The total envelope is a combination of
these, giving a partially-processed elemental abundance that comes from the
dilution of fully processed material by the mass that is still in an unprocessed
form.

With these assumptions, the rate of change of the primordial (1st generation) gas
mass in the cloud core, $M_{\rm gas,1}$, is the increase from stellar debris and
envelope accretion minus what goes into stars,
\begin{eqnarray}
{\dot M}_{\rm gas,1}(t)=\int_0^t \left({{f_{\rm debr}{\dot M}_{\rm star,HM}(t^\prime)}\over{t_{\rm evol}}}
\right)\left(1-f_{\rm p}(t^\prime)\right)\left[1-{{t-t^\prime}\over{t_{\rm evol}}}\right]
dt^\prime\label{md1}\\
 +R_{\rm acc}(t)-(1-f_{\rm p}(t)){\dot M}_{\rm star}(t).
\nonumber
\end{eqnarray}
The first term in the integral is from ejection of stellar debris (equatorial
disks, collisional debris, Roche-lobe overflow). The term $f_{\rm debr}$ is the
average fraction of the mass of a high-mass star that is in the envelope and can be
ejected. It equals $0.651$ for $M_{\rm upper}=100\;M_\odot$ and $0.568$ for $M_{\rm
upper}=300\;M_\odot$ \citep{prantzos06}.  Division by $t_{\rm evol}$ indicates that
we assume this debris is ejected steadily over the evolution time of the massive
star, nominally assumed to be $t_{\rm evol}=3$ Myr. The quantity $f_{\rm
p}(t^\prime)$ is the processed fraction in the gas at time $t^\prime$, and
therefore also the processed fraction in stars that form at time $t^\prime$,
assuming rapid mixing. Thus, $1-f_{\rm p}(t^\prime)$ is the unprocessed fraction of
the mass of the star that previously formed at $t^\prime$. The last term,
$1-([t-t^\prime]/t_{\rm evol})$, tracks the remaining unprocessed fraction in the
stellar envelope as the concentration of processed material increases linearly with
time. This linear increase assumes the p-process elements from the stellar core mix
into the stellar envelope at a steady rate.  In addition to the integral that
represents debris output, the unprocessed gas mass also increases by accretion at
the rate $R_{\rm acc}$. From these additions we subtract the unprocessed gas mass
in the cloud core that goes into stars.

The rate of change of processed gas mass in the cloud core, $M_{\rm gas,2}$ (2nd
generation), is from the addition of stellar debris minus what goes into stars:
\begin{eqnarray}\label{md2}
{\dot M}_{\rm gas,2}(t)=\int_0^t \left({{f_{\rm debr}{\dot M}_{\rm star,HM}(t^\prime)}\over{t_{\rm evol}}}
\right)\left(1-f_{\rm p}(t^\prime)\right)\left({{t-t^\prime}\over{t_{\rm evol}}}\right)dt^\prime\\
\nonumber
+\int_0^t \left({{f_{\rm debr}{\dot M}_{\rm star,HM}(t^\prime)}\over{t_{\rm evol}}}
\right)f_{\rm p}(t^\prime) dt^\prime -f_{\rm p}(t){\dot M}_{\rm star}(t).
\end{eqnarray}
The first integral represents the originally unprocessed mass fraction in the star
when it formed, $(1-f_{\rm p}(t^\prime))$, that came out as debris and became more
and more contaminated with time (as measured by $[t-t^\prime]/t_{\rm evol}$), and
the second integral represents the return of originally processed mass (at fraction
$f_{\rm p}$) into the cloud core. Note that the sum of the processed and
unprocessed gas mass rates from equations (\ref{md1}) and (\ref{md2}) equals
$\int_0^t \left[f_{\rm debr}M_{\rm HM}(t^\prime)/t_{\rm
evol}\right]dt^\prime+R_{\rm acc}-dM_{\rm star}/dt$, which is the total debris rate
plus the accretion rate minus the star formation rate.

Now we determine the masses of primordial and enriched gas in the star-forming
cloud core by integration,
\begin{equation}
M_{\rm gas,1}(t)=\int_0^t {\dot M}_{\rm gas,1}(t)dt
\end{equation}
\begin{equation}
M_{\rm gas,2}(t)=\int_0^t {\dot M}_{\rm gas,2}(t)dt,
\end{equation}
we combine these to get the total gas mass,
\begin{equation}
M_{\rm gas}=M_{\rm gas,1}+M_{\rm gas,2},
\end{equation}
and we determine the mass fractions of enriched gas used above.
\begin{equation}
f_{\rm p}(t)={{M_{\rm gas,2}}\over{M_{\rm gas}}}.
\label{fp}
\end{equation}

The low mass stars do not contribute to the above equations except as a long-term
sink for stellar mass.  However, these stars are important for the observation of
GCs today, and this is where stellar ejection and evaporation come in. We assume
here that only low and intermediate mass stars leave the cluster by these
processes, and we trace only the low mass stellar loss because the intermediate
mass stars will have disappeared by now anyway, except as residual collapsed
remnants. The high mass stars are assumed to segregate to the center of the GC
where essentially all of their p-processes elements are available for gas
contamination, as written in the above equations.  Thus we need to model the escape
of low mass stars.

The discussion in Section \ref{intro} suggests that multi-star interactions and gas
motions in the GC core occasionally accelerate low mass stars up to escape speed or
beyond. Thus the rate of stellar ejection depends on the dynamical rate in the
core, and this is directly proportional to both the free-fall rate and the
consumption rate in the basic model. This implies that there is an additional rate
of change of the mass of low-mass stars, so equation (\ref{e1}) should be revised
to contain an additional term,
\begin{equation}
dM_{\rm star,LM}/dt= f_{\rm LM} M_{\rm gas}(t)/t_{\rm consume}-
f_{\rm eject} M_{\rm star,LM}/t_{\rm consume}
\label{eject}
\end{equation}
for $f_{\rm eject}$ a number less than unity. Recall that $t_{\rm
consume}\sim10t_{\rm ff}$, so the ejection time for $f_{\rm eject}=0.4$ is 25 free
fall times.

\subsection{Analytical solution to a dimensionless model}
\label{analyticalsol}

The time dependence of the star formation rate and cloud mass have analytical
solutions that are conveniently written using the above equations in dimensionless
form. We normalize the gas and stellar masses to the initial cloud core mass,
$M_{\rm gas}(t=0)$, and the time to the consumption time, $t_{\rm consume}$,
assumed to be constant. Normalized quantities are denoted with a tilde. Then the
star formation rate is
\begin{equation}
{{d{\tilde M}_{\rm star}}\over{d{\tilde t}}}={\tilde M}_{\rm gas}.
\label{dless1}
\end{equation}
The gas mass changes from the addition of stellar debris from high mass stars and
cloud envelope accretion and the subtraction of new stars:
\begin{equation}
{{d{\tilde M}_{\rm gas}}\over{d{\tilde t}}}={{f_{\rm debr}f_{\rm HM}}\over{{\tilde t}_{\rm evol}}}
\int_0^{\tilde t} {{{\tilde M}_{\rm star}(t^\prime)}\over{d{\tilde t}}}dt^\prime + {\tilde R}_{\rm acc}
-{{d{\tilde M}_{\rm star}}\over{d{\tilde t}}}.
\label{dless2}
\end{equation}
Here, ${\tilde R}_{\rm acc}=R_{\rm acc}t_{\rm consume}/M_{gas}(t=0)$.

Equation (\ref{dless2}) can be differentiated with respect to time and equation
(\ref{dless1}) can be substituted to give a second order linear differential
equation,
\begin{equation}
{{d^2{\tilde M}_{\rm gas}}\over{d{\tilde t}^2}}+
{{d{\tilde M}_{\rm gas}}\over{d{\tilde t}}}-\gamma{\tilde M}_{\rm gas}=0
\end{equation}
where
\begin{equation}
\gamma={{f_{\rm debr}f_{\rm HM}t_{\rm consume}}\over{t_{\rm evol}}}
\label{secondorder}
\end{equation}
is a dimensionless measure of the relative rate of return of processed gas. For the
numbers in the basic model, $\gamma\sim10^{-2}$. The general solution of equation
(\ref{secondorder}) is the sum of ${\tilde M}_{\rm gas}\propto\exp\left({-\alpha
{\tilde t}}\right)$ with two values of $\alpha$,
\begin{equation}
\alpha=0.5\left(1\pm\left[1+4\gamma\right]^{0.5}\right).
\end{equation}
Using equation (\ref{dless2}) again to fit $d{\tilde M}_{\rm gas}/d{\tilde t}$ at
${\tilde t}=0$, and defining $\Gamma=(1+4\gamma)^{0.5}$, we obtain the solution for
normalized gas mass,
\begin{equation}
{\tilde M}_{\rm gas}({\tilde t})=\left({{\Gamma+0.5-{\tilde R}_{\rm acc}}\over{\Gamma}}\right)
e^{-0.5(1+\Gamma){\tilde t}}
+\left({{\Gamma-0.5+{\tilde R}_{\rm acc}}\over{\Gamma}}\right)e^{0.5(\Gamma-1){\tilde t}}.
\label{total}
\end{equation}
The normalized stellar mass is the time integral over the normalized gas mass, or,
ignoring stellar ejection,
\begin{equation}
{\tilde M}_{\rm stars}({\tilde t})=
\left({{\Gamma+0.5-{\tilde R}_{\rm acc}}\over{0.5\Gamma(1+\Gamma)}}\right)
\left(1-e^{-0.5(1+\Gamma){\tilde t}}\right)
+\left({{\Gamma-0.5+{\tilde R}_{\rm acc}}\over{0.5\Gamma(\Gamma-1)}}\right)
\left(e^{0.5(\Gamma-1){\tilde t}}-1\right).
\end{equation}
This solution consists of an exponentially decaying initial burst of star formation
and gas depletion with a timescale approximately equal to the consumption time
($0.5(1+\Gamma)\sim1$ in normalized units), followed by a slowly growing gas mass
and star formation rate as accretion and stellar debris add to the cloud core (with
growth rate $0.5(\Gamma-1)\sim\gamma<<1)$.

\section{Results}
\label{results}

\begin{figure}\epsscale{.8}
\plotone{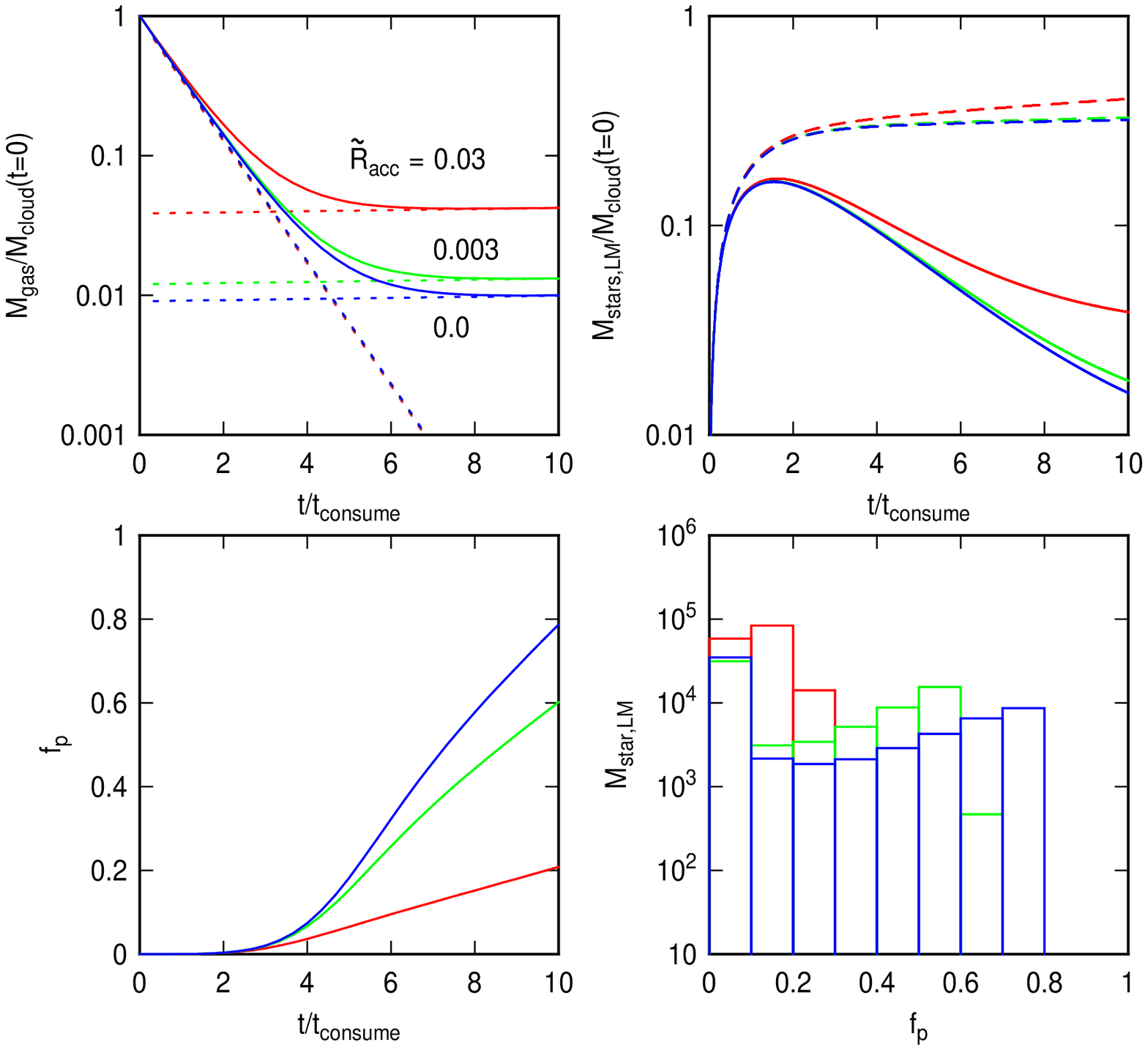}
\caption{Models for the formation of globular clusters with three different normalized
accretion rates onto the cloud core. The top-left panel shows the relative core gas mass
as a function of time, in units of the gas consumption time. The dotted lines are the
separate exponential solutions from equation (\ref{total}), whose sum equals the total
solution. The core mass decreases faster when there is no accretion. The top right panel
shows the mass in low mass stars for these three cases (designated by the corresponding
colors), with dashed lines representing the case when there is no stellar ejection from
the cluster, and solid lines showing the case with ejection using an efficiency of
$f_{\rm eject}=0.4$. The maximum time on the abscissa corresponds to the lifetime of a
massive star, since we assume $t_{\rm consume}/t_{\rm evol}=0.1$ for these results.
Thus the final mass in low mass stars is about one-tenth of the total formed mass in stars of
this type.  The lower left panel shows the time-increase in the processed fraction of gas
in the cloud core, defined as the ratio of the p-processed mass to the current core gas mass.
The processed fraction increases with time as massive stars evolve, and it is higher when the
accretion rate is low because then there is less dilution of stellar envelope material with
pristine gas. The lower right panel shows the distribution of processed fraction among
the low mass stars at the end of the star formation, when $t/t_{\rm consume}=10$.  There is
a continuum of $f_{\rm p}$ because stars form continuously as the core gas is contaminated.
We consider the first generation stars to be those in the lowest bin.  The numbers of these
stars is relatively large in all cases, but the cumulative number in bins of higher $f_{\rm p}$ can be
larger.} \label{walch16sfr2gasmass}\end{figure}

Figure \ref{walch16sfr2gasmass} shows numerical solutions to equations (\ref{e1})
to (\ref{fp}) with mass and time normalized as above. The three curves are for
different normalized accretion rates, as indicated by their colors and the labels
for ${\tilde R}_{\rm acc}$. The solutions have the parameters discussed above for
an IMF extending out to $300\;M_\odot$, i.e., $f_{\rm HM}=0.164$ and $f_{\rm
debr}=0.568$, and they assume $t_{\rm consume}/t_{\rm evol}=0.1$, which together
make $\gamma=0.0093$. The top left panel shows the cloud core gas mass as a
function of time and the top right panel shows the mass of low mass stars without
(dashed curves) and with (solid curves) ejection assuming $f_{\rm eject}=0.4$.  The
lower left panel shows the processed fraction in the cloud core, which is also the
processed fraction in the stars that form at that time. The analytical solutions to
the gas and stellar masses fit exactly on top of the numerical solutions and are
not shown in the figure. However, the top left panel has each exponential function
from the analytical solution plotted separately as a dotted line; the total
solution is the sum of these, as in equation (\ref{total}).

The lower right panel shows the distribution of the processed fraction, $f_{\rm
p}$, among low-mass stars after star formation is assumed to stop, which is at the
time $t_{\rm evol}$ in these models to avoid supernova contamination. All three
cases assume stellar ejection with $f_{\rm eject}=0.4$. Recall that in observed
GCs, $f_{\rm p}$ can range up to 0.7 from the Li abundance \citep{dec07b}, and that
is how far the blue histogram goes (which is for ${\tilde R}_{\rm acc}=0$). Also
from observations, the ratio of stellar mass in the second generation to total
stellar mass ranges from $\sim0.5$ to $\sim0.8$. For the 3 histograms in Figure
\ref{walch16sfr2gasmass}, the ratios of the masses in all but the lowest-$f_{\rm
p}$ interval (i.e., the G2 stars) to the sum of all the masses (the G1+G2 stars) is
about the same as the observations: 0.63 for the red curve (${\tilde R}_{\rm
acc}=0.03$), 0.50 for the green curve (${\tilde R}_{\rm acc}=0.003$) and 0.43 for
the blue curve (${\tilde R}_{\rm acc}=0$).  The processed fraction $f_{\rm p}$
increases with time as more and more massive stellar envelopes with their
ever-increasing p-process contaminations disperse inside the cloud core (lower left
panel). The maximum value of $f_{\rm p}$ at the end of the star formation time
(right-hand limit to the curves in the lower-left panel) increases with decreasing
${\tilde R}_{\rm acc}$ because there is less dilution of the cloud core gas with
pristine infall from the cloud envelope.

In equation (\ref{eject}), ejection operates on all low-mass stars regardless of
when they form, but it tends to remove a higher proportion of the first stars that
form, i.e., weighted toward the G1 stars, than the later stars that form because
the first stars have been exposed for the longest time to the multi-star
interactions and gas motions that cause ejection. The top right panel of Figure
\ref{walch16sfr2gasmass} shows that ejection with the assumed $f_{\rm eject}=0.4$
reduces the mass in low mass stars by about a factor of 10. Without such stellar
loss, the relative proportion of G2 stars shown in the lower right would be much
less; i.e., the lowest $f_{\rm p}$ bin would be much higher compared to the others,
although the $f_{\rm p}$ range in the histogram would be the same.

Related to this age-dependence of stellar ejection is a prediction from this model
that the G1 stars should on average be less concentrated in the GC than the G2
stars. This is because the G1 stars have had more opportunities to absorb kinetic
energy from multi-stellar interactions in the core than the G2 stars, regardless of
whether the stars escape the cluster.  Such a central concentration of G2 stars is
observed \citep{gratton12}.

\begin{figure}\epsscale{.9}
\plotone{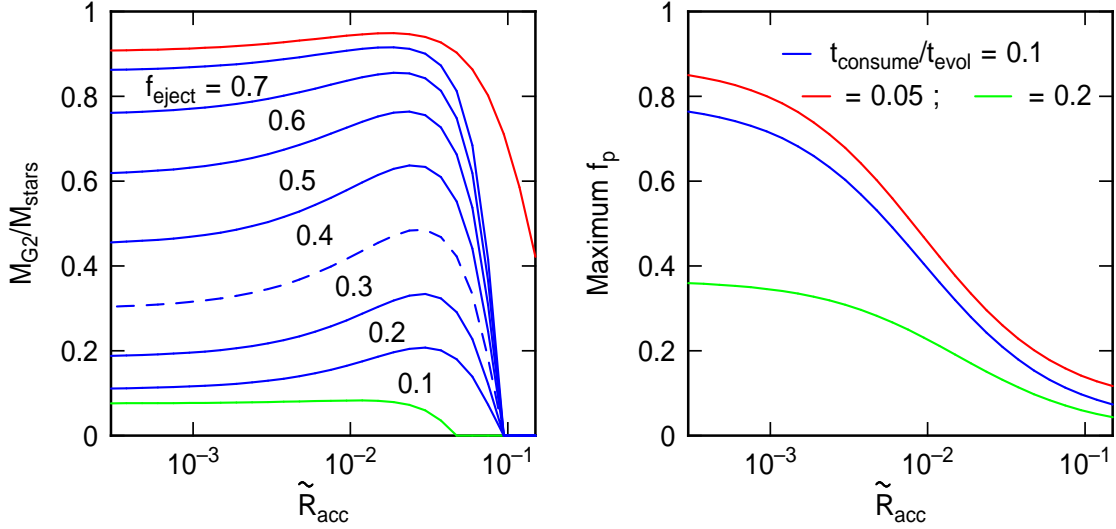}
\caption{(Left) The fraction of low mass stars in the ``2nd generation,''
defined to be those in all but the lowest bin of $f_{\rm p}$ from figure 1.
This fraction is shown as a function of the normalized accretion rate onto the
cloud core, and for different values of the efficiency of stellar ejection from
the cluster, $f_{\rm eject}$. The G2 fraction is low at low ${\tilde R}_{\rm acc}$
because then star formation ends quickly and there is too little time for p-processed
elements to get into the gas for subsequent generations of star formation. This fraction
is low also at high ${\tilde R}_{\rm acc}$ because then the core is highly
diluted with accreted pristine gas. It peaks at around ${\tilde R}_{\rm acc}\sim0.3$,
which is when the accretion rate is 0.3 times the initial core mass divided by the
gas consumption time.  Higher $f_{\rm eject}$ produces higher G2 fractions because then
a higher proportion of early-forming low-mass stars, which have near-G1 abundances, have been
ejected. (Right) The maximum value of the processed fraction, $f_{\rm p}$, is shown
versus the normalized accretion rate for three different cases of the relative consumption time.
The low and high cases are also shown in the left-hand panel using the same colors.
The curves in the right-hand panel are independent of $f_{\rm eject}$, which only
affects the proportion of stars at various $f_{\rm p}$, but not the maximum in $f_{\rm p}$.
The result indicates that G2 stars reach a higher p-processed fraction if the
accretion rate is low, because then the dilution of stellar debris by pristine gas is
smaller.
} \label{walch16g2g1}\end{figure}

The left-hand panel of Figure \ref{walch16g2g1} shows the effect of stellar
ejection through the parameter $f_{\rm eject}$ on the fraction of the mass in the
G2 population.  The blue curves assume the usual $t_{\rm consume}/t_{\rm evol}=0.1$
with a range of $f_{\rm eject}$ in equal steps between 0.1 and 0.7. The dashed blue
curve with $f_{\rm eject}=0.3$ is one of this sequence but it is highlighted to
contrast it with the red and green curves, which assume $f_{\rm eject}=0.3$ but
also $t_{\rm consume}/t_{\rm evol}=0.05$ and 0.2, respectively, for the same value
of $\gamma=f_{\rm debr}f_{\rm HM}t_{\rm consume}/t_{\rm evol}$ (i.e., $f_{\rm
debr}$ is $2\times$ larger and smaller when $t_{\rm evol}$ is $2\times$ larger and
smaller, respectively, to keep $\gamma$ the same). There are three important
dependencies shown in this figure: (1) the G2 mass fraction is low for both low
accretion and high accretion rates, with a peak at ${\tilde R}_{\rm acc}\sim0.03$;
(2) the G2 mass fraction increases with higher ejection parameter, and (3) the G2
mass fraction depends strongly on the ratio of the gas consumption time to the
stellar evolution time.

The first result arises because high ${\tilde R}_{\rm acc}$ causes severe dilution
of the p-processed material for later stars that form, and because low ${\tilde
R}_{\rm acc}$ causes most of the stars to form quickly and deplete the gas before
p-processed elements have time to get into the cloud core. The second result is a
consequence of ejecting more early-forming stars than late-forming stars because of
the greater exposure of early-forming stars to time-changing gravitational forces.

The third result indicates the importance of the dimensionless parameter $t_{\rm
consume}/t_{\rm evol}$. A low value means there are more dynamical times available
for stellar ejection before the supernova era begins (at $t=t_{\rm evol}$), so the
first-forming stars become much more depleted compared to the last-forming stars.
This increases $M_{\rm G2}/M_{\rm stars}$, as shown on the left of Figure
\ref{walch16g2g1}. It also means there are more massive stars at early times
compared to late times, because of the more rapid drop in the star formation rate
from faster gas consumption at small $t_{\rm consume}$. These greater numbers of
early massive stars send relatively more p-processed matter into the smaller amount
of remaining gas mass in the cloud core, increasing the maximum $f_{\rm p}$. These
trends with lower $t_{\rm consume}/t_{\rm evol}$ occur when the density of the
cloud core increases, because that lowers the free fall time and, correspondingly,
$t_{\rm consume}$, at a fixed stellar evolution time. Thus we have the important
result that {\it p-process contamination is larger and involves a higher fraction
of remaining long-lived stars in a globular cluster when the initial cluster gas
density is higher.}  This may be the critical clue to the distinct origin of old
globular clusters.

Figure \ref{walch16g2g1} suggests that realistic results require a fairly high
ejection parameter $f_{\rm eject}$, on the order of several tenths, which
corresponds to a mean time before ejection of low mass stars equal to several tens
of dynamical times ($=t_{\rm consume}/(t_{\rm ff}f_{\rm eject})$).  \cite{fujii13}
follow the ejection of stars of various masses from the collapsed core of a cluster
owing to binary star interactions. The number of ejected high mass stars per
cluster is fairly low, although it can account well enough for runaway OB stars in
the field. The number of ejected low mass stars, which are of greatest interest in
the present context, is much higher, following the stellar IMF (see their figure
1). They did not do a simulation that is exactly what we need, which would involve
a lowest stellar mass less than $0.1\;M_\odot$ and a cluster more massive than
$10^6\;M_\odot$. In their most relevant case, which had the lowest mass stars, the
fraction of stars ejected was 0.4\%. That is only for the binary star mechanism,
however. Here we envision a much more dynamic environment in which gaseous mass
motions from multi-star interactions and winds stochastically change the whole
cluster core potential, as in simulations of dwarf galaxy nuclei \citep{elbadry16}.
That second process, in addition to the possible presence of several dense cores in
a proto-GC, each of which has one or more important massive binaries for stellar
ejection, could possibly bring the ejection fraction per dynamical time up to the
required value.

\begin{figure}\epsscale{.9}
\plotone{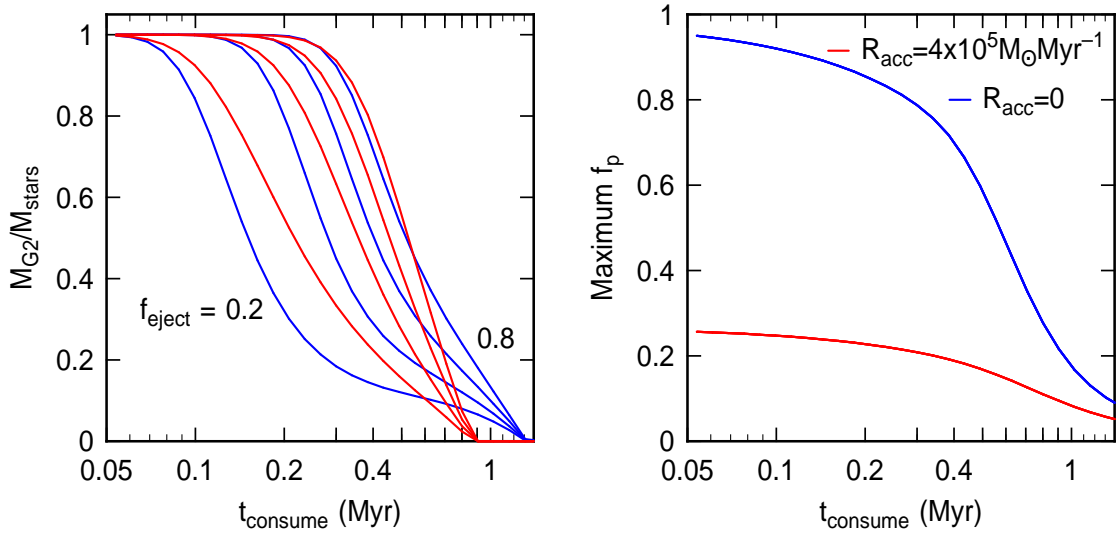}
\caption{The effect is shown of the star formation consumption time on the
G2 mass fraction (left) and the maximum processed fraction (right).
Longer consumption times correspond to slower star formation and less
production of p-processed gas before the supernova era begins. They also
correspond to less stellar ejection from the cluster, which should
operate on the dynamical time (proportional to the consumption time)
and therefore to a greater residual of G1 stars, also lowering the
G2 fraction.} \label{walch16consume}\end{figure}

The dependence of relative G2 mass and processed fraction, $f_{\rm p}$, on $t_{\rm
consume}$ is shown in Figure \ref{walch16consume}, where we assume fixed $f_{\rm
debr}$, $f_{\rm HM}$, $f_{\rm eject}$, and $t_{\rm evol}$ and vary only $t_{\rm
consume}$.  In this case the input parameters are dimensional, with $t_{\rm
consume}$ up to $\sim1.4$ Myr and $R_{\rm acc}=0$ (blue curves) and
$4\times10^5\;M_\odot$ Myr$^{-1}$ (red curves). There are sharp decreases in the
relative mass of the second generation (left panel) and the maximum processed
fraction (right panel) as the consumption time increases, which corresponds to a
decrease in cloud core density. Because the bulk peculiarity of old GCs is in these
two quantities, it seems evident that {\it the primary distinction between clusters
that formed in the early universe and most of the main-disk clusters forming today
is cloud-core density}.

% changed per yr to per Myr

\section{Conclusions}
\label{conclu}

We present a model of star formation in massive dense clusters that may be relevant
to the formation of old globular clusters. This model has several key features that
should cause a cluster to form stars with a wide range of p-processed elements, in
agreement with observations. These features are:
\begin{itemize}
\item A $10^6-10^7\;M_\odot$ cloud core with a density of $\sim 10^6$
    cm$^{-3}$, giving a free fall time of $\sim0.03$ Myr. The corresponding
    virial velocity is $\sim 80$ km s$^{-1}$, which is comparable to the
    turbulent speed in an L* galaxy at the same redshift where the GC forms.
    Such a core is tightly bound and should form stars with an elevated star
    formation efficiency per unit free fall time, assumed here to be
    $\sim10$\%. The free fall time and efficiency combine to give a gas
    consumption time that is much shorter than the evolution time of a high
    mass star. Then many generations of stars form in the gaseous debris of
    earlier generations.

\item Interactions between massive stars at close range, including massive
    binary star interactions with single stars and massive stellar mergers,
    which lead to the swelling and dispersal of massive stellar envelopes and
    the dispersal of extruded equatorial disks around rapidly rotating massive
    stars. The dispersed envelope gas carries p-processed elements into the
    cloud core where it gets incorporated into new stars.

\item Rapid stellar ejection driven by the time-changing gravitational
    potential of multi-star interactions and pressurized gas motions. The
    ejection rate is assumed to scale with the dynamical rate in the cluster.
    We point out that stars ejected from a cluster with a terminal velocity
    equal to only several tenths of the cloud core virial speed can move
    through the host galaxy at greater than the galaxy's escape speed and
    thereby leave the host altogether. Such stellar loss may solve the problem
    of missing halo stars in Fornax and WLM, which have very low escape speeds.

\item Star formation in the cloud core using gas that is a combination of
    original core gas, stellar debris that becomes more and more contaminated
    by p-processed elements over time, and newly accreted gas with the original
    abundances. Because the consumption time at high density is much less than
    the lifetime of a massive star, many generations of stars form before
    supernovae finally clear the gas away. We assume that each generation has a
    complete and normal IMF. We track the high mass stars for the production of
    p-processed elements, and the low mass stars for comparison to GCs today.
\end{itemize}

As a result of these assumptions, we produce GC stellar populations with
approximately the observed range, mass, and radial distributions of p-process
contamination.  We find that the mass fraction of the final GC in the form of
contaminated stars (i.e., the ``2nd generation'' fraction), and the maximum amount
of p-process contamination in the late-forming stars, both increase strongly with
cloud core density.  This result suggests that the primary difference between old
GCs with their significant amounts of p-process contamination and young
super-massive star clusters that show little evidence for self-enrichment is that
the cluster-forming clouds at high redshift have higher densities.  This is to be
expected because high redshift galaxies have larger gas fractions, higher gas
surface densities and faster turbulent speeds than all but the most active galaxies
today. As a result, the pressures in high-redshift galaxies are high, and the
densities in the star-forming cloud cores should be high too.

This paper benefited from interesting discussions with Dr. Stefanie Walch at an
early stage of this investigation, and from comments by the referee.

\clearpage
%fig1

\clearpage
%fig2

\clearpage
%fig3

\end{document}